\documentclass{article}

\usepackage{arxiv}

\usepackage[utf8]{inputenc} 
\usepackage[T1]{fontenc}    
\usepackage{hyperref}       
\usepackage{url}            
\usepackage{booktabs}       
\usepackage{amsfonts}       
\usepackage{nicefrac}       
\usepackage{microtype}      
\usepackage{lipsum}		
\usepackage{graphicx}
\usepackage{amsmath,amssymb,amsfonts,subcaption} 
\usepackage{natbib}
\usepackage{doi}
\usepackage{csquotes}

\title{How the Graph Construction Technique Shapes Performance in IoT Botnet Detection: Insights from Graph Attention Networks}

\author{ \href{https://orcid.org/0000-0002-6972-8935}{\includegraphics[scale=0.06]{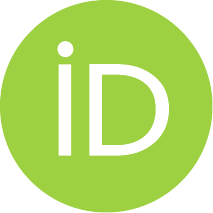}\hspace{1mm}Hassan~Wasswa}
        \\
	School of Systems and Computing\\
	University of New South Wales\\
	Canberra, 2600 ACT Australia \\
	\texttt{h.wasswa@unsw.edu.au} \\
	\And
	\href{https://orcid.org/0000-0002-8837-0748}{\includegraphics[scale=0.06]{orcid.pdf}\hspace{1mm}Hussein~Abbass} \\
	School of Systems and Computing\\
	University of New South Wales\\
	Canberra, 2600 ACT Australia \\
	\texttt{h.abbass@unsw.edu.au} \\
    \And
	\href{https://orcid.org/0000-0001-7934-5658}{\includegraphics[scale=0.06]{orcid.pdf}\hspace{1mm}Timothy~Lynar} \\
	School of Systems and Computing\\
	University of New South Wales\\
	Canberra, 2600 ACT Australia \\
	\texttt{t.lynar@unsw.edu.au} \\
}

\renewcommand{\shorttitle}{How the Graph Construction Technique Shapes Performance in IoT Botnet Detection}

\hypersetup{
pdftitle={A template for the arxiv style},
pdfsubject={q-bio.NC, q-bio.QM},
pdfauthor={David S.~Hippocampus, Elias D.~Striatum},
pdfkeywords={First keyword, Second keyword, More},
}

\begin{document}
\maketitle

\begin{abstract}
	The increasing incidence of IoT-based botnet attacks has driven interest in advanced learning models for detection. Recent efforts have focused on leveraging attention mechanisms to model long-range feature dependencies and Graph Neural Networks (GNNs) to capture relationships between data instances. Since GNNs require graph-structured input, tabular NetFlow data must be transformed accordingly. This study evaluates how the choice of the method for constructing the graph-structured dataset impacts the classification performance of a GNN model. Five methods---k-Nearest Neighbors, Mutual Nearest Neighbors, Shared Nearest Neighbor, Gabriel Graph, and $\epsilon$-radius Graph---were evaluated in this research. To reduce the computational burden associated with high-dimensional data, a Variational Autoencoder (VAE) is employed to project the original features into a lower-dimensional latent space prior to graph generation. Subsequently, a Graph Attention Network (GAT) is trained on each graph to classify traffic in the N-BaIoT dataset into three categories: Normal, Mirai, and Gafgyt. The results indicate that using Gabriel graph achieves the highest detection performance with an accuracy of 97.56\% while SNN recorded the lowest performance with an accuracy as low as 78.56\%.
\end{abstract}


\section{Introduction}
Given the rising incidence of IoT-based botnet attacks, enhancing the effectiveness of learning models for detecting such threats has emerged as a key priority for researchers and practitioners in the field of IoT security. Prior research has examined a range of techniques, including conventional machine learning methods~\cite{ali2024hybrid,bojarajulu2023intelligent,wasswa2022proof,10063566}, sophisticated deep learning architectures~\cite{pavithran2025iot, sana2024securing, wasswa2023enhancing}, and hybrid models aimed at strengthening network security, with a particular emphasis on IoT botnet detection~\cite{sinha2025high, ali2024hybrid, bojarajulu2023intelligent}. A recent and increasingly prominent approach involves the integration of attention mechanisms, which are designed to capture long-range dependencies within input features. Notably, studies such as~\cite{latent_dim_impact10511431, WASSWA2025127504, vit_classifier_10426522} have demonstrated that attention-based models can achieve near-optimal performance in the detection of IoT botnet traffic.

Nevertheless, conventional models often treat each attack instance as an isolated data point, overlooking the potential inter-dependencies among different attack instances. To overcome this limitation, recent research has increasingly turned to exploring Graph Neural Network (GNN)-based approaches for attack detection~\cite{ahanger2025advanced, lin2025gracl, do2025investigating, altaf2024gnn, protogerou2021graph}. These models are trained to generate an embedding space in which nodes with similar characteristics---and their relationships to neighboring nodes---are positioned in close proximity. To further boost detection performance, advanced methods have incorporated attention mechanisms into GNN architectures. This hybrid design enables the model to capture both long-range feature dependencies, through attention, and local inter-instance relationships, via graph structure, leading to more robust and accurate detection of IoT botnet traffic.

Since the feature extraction process from the NetFlow capture (.pcaps) often results into tabular datasets (such as the .csv format), converting the dataset into a graph structure is a common preprocessing step when applying GNNs to network traffic classification tasks. The method used to construct the graph---i.e., how nodes (samples) are connected via edges---plays a critical role in shaping the graph's topology and, consequently, the model's performance. However, the impact of different graph construction strategies on the classification performance of GNN models remains an open question in the existing literature. 

To tackle this challenge, this work evaluates five .csv dataset-to-graph-dataset construction techniques---k-nearest neighbors (kNN), epsilon-radius graph (\(\epsilon\)-graph), mutual nearest neighbors (MNN), Gabriel-graph, Shared nearest neighbors (SNN)---and evaluates the impact of each method on the performance of a Graph Attention Neural Network (GAT) model for botnet attack detection on the N-BaIoT dataset.

The rest of this paper is organized as follows.
Section~\ref{related_work} presents prior work integrating the GNN model with the attention mechanism for anomaly detection. This is followed by section~\ref{methodology} where a detailed description of the framework used in this work is presented, together with a brief description of the dataset and the various techniques used. In section~\ref{results} we present and discuss the findings of this study followed by the conclusion in section~\ref{conclusion}.

\section{Related Work}\label{related_work}
Recent research has increasingly explored hybrid models that combine attention-based transformer architectures with GNNs to enhance anomaly detection capabilities. These hybrid approaches capitalize on the complementary strengths of GNNs, which are adept at capturing complex node-edge relationships, and transformers, known for their effectiveness in modeling long-range dependencies across features---resulting in superior detection performance.

For example, the authors of study~\cite{ahanger2025advanced} aimed to address IoT security challenges caused by heterogeneous interconnected devices, where conventional methods such as encryption and authentication often prove inadequate. The authors proposed a graph-based approach, utilizing a Graph Attention Network (GAT)-driven Intrusion Detection System (IDS) to analyze IoT graphs. By evaluating the method on the NSL-KDD dataset, the study demonstrated the potential of graph neural networks as a scalable and robust solution for intrusion detection in IoT systems.

 The study in~\cite{zhang2024hybrid} introduced a hybrid model integrating a GNN with a transformer-based architecture specifically for intrusion detection. In this model, the GNN component is designed to capture intricate structural dependencies among nodes and edges, while the transformer module improves anomaly detection by effectively modeling extended contextual relationships.

In a related development, \cite{kumar2025grma} presented a framework that incorporates a graph convolutional neural network and an attention mechanism to analyze graph-structured data from IoT devices. This integration facilitates the detection of botnets by accurately modeling device interactions and uncovering latent patterns indicative of malicious behavior.


Study~\cite{wasswa2025graph} investigated the influence of three dimensionality reduction methods---AE-encoder, VAE-encoder, and PCA—on the performance of a Graph Attention Network (GAT) for IoT botnet detection. In their experiments, the original 84-dimensional CICIoT2022 feature space was compressed into 8 dimensions before being converted into graph-structured data for training. The results demonstrated that VAE-encoder yielded the best performance among the reduction methods. Additionally, the $3\text{-euclidean}$ \enquote{n\_neighbors-metric} configuration showed superior results in graph construction using the kNN algorithm compared to other parameter combinations.

The authors of study~\cite{zhang2023malicious} aimed to address the challenge of detecting complex IoT cyberattacks by introducing EGAT-LSTM, a model that integrates an enhanced Graph Attention Network with an LSTM. The approach captures spatial and temporal characteristics of network traffic while refining feature aggregation to retain essential information. The study applied this method to IoT malicious traffic classification, offering a more effective solution than conventional single-flow analysis.

The work in~\cite{wasswa2025gnns} compared the classification performance of GNN-based models—VAE-GCN and VAE-GAT—with two advanced deep learning models---VAE-MLP and ViT-MLP—using the 115-dimensional N-BaIoT dataset reduced to 8 dimensions. The findings showed that VAE-GAT outperformed VAE-GCN; however, both GNN-based models were generally outperformed by the other deep learning architectures, indicating a need for further refinement of GNN models to achieve competitive classification results.

In another significant contribution, \cite{li2025multi} focused on detecting cyber-attacks targeting electric vehicle (EV) charging infrastructure. Their methodology involved creating graph representations from hardware logs and applying a GNN to model the interrelationships among features. The inclusion of a transformer further enhanced the model’s ability to capture intricate feature interactions, thereby improving the accuracy of attack detection.

Additionally, \cite{xu2025ajsage} proposed AJSAGE, a model that extends the GNN-based GraphSAGE architecture with an attention mechanism. This augmentation enhances the model’s effectiveness in identifying anomalous traffic nodes within graph-based network attack data, particularly in detecting sophisticated cyber threats.

Collectively, these studies highlight a growing momentum toward hybrid GNN-transformer architectures in anomaly detection tasks, underlining their promising applicability in areas such as intrusion detection, botnet analysis, and IoT cybersecurity.

\section{Methodology}\label{methodology}
\begin{figure}[!ht]
     \centering
    \includegraphics[scale=1.0]{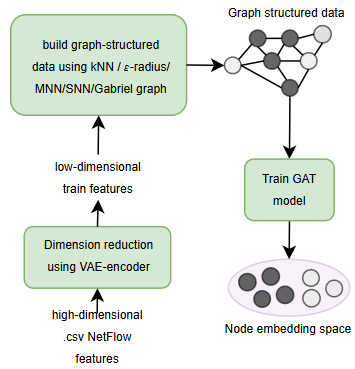}
    \caption{Detection framework}
    \label{fig:framework}
\end{figure}

This study aimed to examine the impact of different graph data construction techniques to the IoT botnet detection performance of the Graph Attention (GAT) neural network model. The study deploys the architecture proposed in study~\cite{wasswa2025graph, wasswa2025gnns} which involves transforming the high-dimensional .csv NetFlow features to a low-dimensional representation. Since study~\cite{wasswa2025graph} revealed that the VAE-encoder-based dimension reduction method outperforms other reduction methods including PCA and the classical AE-encoder, this study deployed the VAE-encoder for dimension reduction transforming the 115-dimensional dataset into a 6-dimensional representation. Each of the five graph construction techniques is used in turn to convert the low-dimensional dataset into a graph structured dataset that can be processed by the GAT model. The architectural framework utilized in this study is illustrated in Figure~\ref{fig:framework}. A detailed description of the various techniques deployed in this study is provided in the subsequent subsections.

\subsection{Dataset}\label{subsec:dataset}
The models were evaluated using the N-BaIoT dataset~\cite{meidan2018n}, which includes NetFlow data from nine IoT devices infected with \enquote{Mirai} and \enquote{Gafgyt} malware. The classification task involved distinguishing between \enquote{Normal}, \enquote{Mirai}, and \enquote{Gafgyt} traffic. After removing duplicates, 2,482,470 instances remained. To address class imbalance, the dataset was down-sampled to {\enquote{Normal}: 500,000 (40.58\%), \enquote{Mirai}: 500,000 (40.58\%), \enquote{Gafgyt}: 232,258 (18.84\%)} for training and evaluation.

\subsection{Variational Autoencoder}
Variational Autoencoders (VAEs)~\cite{kingma2019introduction} address the lack of latent space regularization in classical AEs by employing Bayesian variational inference. A VAE introduces a latent variable $z$ to approximate the joint distribution $p_{\psi}(x, z)$ and estimate the marginal $p_{\psi}(x)$. Assuming a simple prior for $z$, and approximating $p_{\psi}(z|x)$ with $q_{\beta}(z|x)$, it optimizes parameters $\beta$ by maximizing the expected log-likelihood:

\begin{equation} \label{eqn1}
E_{z \sim q_{\beta} (z \vert x)}\log p_{\psi}(x) = \mathcal{L}(\psi, \beta, z) + D_{KL}(q_{\beta}(z \vert x) \vert\vert p_{\psi}(z \vert x))
\end{equation}

Here, $\mathcal{L}$ is the ELBO and $D_{KL}$ is the divergence between approximate and true posteriors. Training maximizes ELBO and minimizes $D_{KL}$.

\subsection{k-Nearest Neighbors (kNN) Graph Construction}\label{subsec:k_NN}
The k-Nearest Neighbors (kNN) graph~\cite{zhu2003semi} is a widely used method for converting tabular data into graph structures. In this technique, each data point is treated as a node and connected to its $k$ nearest neighbors based on a distance metric, typically the Euclidean distance, defined in eq.~\ref{eqn2}:

\begin{equation} \label{eqn2}
d(x_i, x_j) = \sqrt{\sum_{l=1}^d (x_i^{(l)} - x_j^{(l)})^2}
\end{equation}

Given a dataset $X = {x_1, x_2, \ldots, x_n}$ with $x_i \in \mathbb{R}^d$, edges are formed from each node $x_i$ to its $k$ closest nodes, resulting in a directed graph. This can be made undirected by symmetrization. The kNN graph is appreciated for its simplicity and ability to preserve local geometric relationships, preventing isolated nodes. However, its effectiveness is highly sensitive to the choice of $k$: small values may lead to disconnected graphs, while large values risk introducing noise, especially in high-dimensional data. Despite these limitations, kNN remains a fundamental tool in graph-based learning due to its ease of implementation and practical utility.

\subsection{Mutual Nearest Neighbors (MNN) Graph}\label{subsec:mutual_NN}
The Mutual Nearest Neighbors (MNN) graph~\cite{haghverdi2018batch} enhances the standard k-Nearest Neighbors (k-NN) graph by requiring mutual similarity for edge creation. An undirected edge between nodes $i$ and $j$ is added only if $i \in \mathcal{N}_k(j)$ and $j \in \mathcal{N}_k(i)$, where $\mathcal{N}_k(i)$ is the set of $k$-nearest neighbors of node $i$. This reciprocal condition reduces noisy or spurious connections by ensuring edges represent strong, symmetric relationships. As a result, MNN graphs are more robust and better suited for tasks like clustering and anomaly detection, where precision in graph structure is critical. However, the conservative nature of MNN can lead to sparsity or disconnected components, especially in unevenly distributed datasets. Additionally, meaningful but asymmetric relationships may be excluded, potentially affecting performance in certain learning tasks.

\subsection{Shared Nearest Neighbors (SNN)}\label{subsec:shared_nn}
In a Shared Nearest Neighbors (SNN) graph~\cite{ertoz2003finding}, two nodes $A$ and $B$ are connected if they have at least $\theta$ common neighbors in their respective $k$-nearest neighbor sets $\mathcal{N}_k(A)$ and $\mathcal{N}_k(B)$. The similarity between $A$ and $B$ is quantified as in eq.~\ref{eqn3}:

\begin{equation} \label{eqn3}
\text{SNN}(A, B) = |\mathcal{N}\_k(A) \cap \mathcal{N}\_k(B)|
\end{equation}

An edge is added if $\text{SNN}(A, B) \geq \theta$, with the score optionally used as an edge weight. Compared to traditional $k$-NN, SNN better captures local density and structural similarity, making it effective in noisy data and for community detection. It is also more robust to outliers, as isolated nodes rarely share neighbors. However, it is computationally intensive on large datasets due to pairwise comparisons and requires careful selection of the threshold $\theta$, which is often dataset-specific and must be empirically tuned.

\subsection{\(\varepsilon\)-Radius Graph}\label{subsec:epsilon_radius}

The $\varepsilon$-Radius graph~\cite{belkin2003laplacian} is a proximity-based graph where an undirected edge connects two nodes if their distance is less than a predefined threshold $\varepsilon$. For data points $\{x_1, x_2, \dots, x_n\}$ in a metric space with distance function $d(\cdot,\cdot)$, an edge is formed between $x_i$ and $x_j$ if $
d(x_i, x_j) < \varepsilon.
$

This approach preserves the geometric locality of the data and is effective for capturing spatial proximity and density-based relationships. It adapts to local data distributions by forming connections within a fixed radius. However, its performance heavily depends on the choice of $\varepsilon$: a small value may produce a sparse, disconnected graph, while a large one may yield an overly dense and noisy graph. Moreover, it does not guarantee connectivity for all nodes and lacks scale invariance, limiting its robustness on datasets with varying densities.

\subsection{Gabriel Graph}\label{subsec:gabriel_radius}
The Gabriel graph~\cite{gabriel1969new} is a geometric proximity graph where an edge exists between nodes $A$ and $B$ only if no other node $C$ lies within the closed disc whose diameter is $AB$. This condition is defined using the expression in eq.~\ref{eqn4} ensuring local emptiness between connected nodes.

\begin{equation} \label{eqn4}
\|C - \frac{A + B}{2}\|^2 \geq \left(\frac{\|A - B\|}{2}\right)^2, \quad \forall C \neq A, B
\end{equation} 

 Gabriel graphs preserve geometric and topological structure, yielding sparse and planar representations ideal for spatial analysis and visualization. Despite these advantages, they are highly sensitive to noise and small positional changes, which can destabilize the graph. Additionally, their computational cost grows rapidly with dataset size due to the exhaustive pairwise emptiness check. In high-dimensional spaces, the resulting graphs often become overly sparse or under-connected, limiting their effectiveness in tasks that rely on dense graph connectivity for learning performance.

\subsection{Graph Neural Networks}
Traditional deep learning models like CNNs and RNNs perform well on structured data but falter with graph-structured inputs. Graph Neural Networks (GNNs), introduced in~\cite{scarselli2008graph}, overcome this by modeling data as nodes and edges, capturing complex relationships. This makes GNNs particularly suitable for analyzing interconnected systems such as molecular structures, materials, and network flows, offering deeper insights into relational data~\cite{ying2019gnnexplainer, battaglia2018relational, xu2018powerful, hamilton2017inductive, zhou2020graph}.

In NetFlow analysis, graph nodes correspond to individual traffic instances, while edges capture relationships such as cosine similarity, Euclidean distance, or shared IP addresses. Through iterative message passing, a GNN processes the features of each node along with those of its neighbors to construct a rich representation of the traffic behavior. It encodes each node into a vector embedding, placing similar instances nearby in the embedding space. These embeddings facilitate tasks like node classification, allowing the GNN to identify attack types in previously unseen traffic.

\subsection{The attention mechanism}
Recent progress in computer vision, particularly in object detection~\cite{fan2020few, li2020object}, has driven the incorporation of attention mechanisms---especially multi-head attention~\cite{vaswani2017attention}---into NetFlow-based attack detection models~\cite{latent_dim_impact10511431, WASSWA2025127504, vit_classifier_10426522, gogoi2023dga, lu2022self}. Attention mechanisms enhance model performance by focusing on the most relevant input regions through a query-key-value structure: the query ($Q$) seeks specific features, keys ($K$) indicate potentially matching inputs, and values ($V$) hold data related to keys ($K$). The output is a weighted sum of $V$, based on the similarity between $Q$ and $K$.

\subsection{Model Training and evaluation}
The dataset used for training and evaluation was partitioned randomly into training and testing subsets in a $4:1$ ratio. Additionally, during each training epoch, a $0.1$ fraction of the training batch was reserved for validation purposes. For dimension reduction, the VAE model was trained for $20$ epochs on the experimental dataset and using its encoder component, the original high-dimensional training data was transformed into a $6$-dimensional space. For the GAT model, ReLU served as the activation function for neurons, and the Adam optimizer was applied with a learning rate of $0.01$ and a weight decay of $5e^{-4}$. The model was trained for $100$ epochs per graph construction approach—using a batch size of $128$. For the $\varepsilon$-radius graph, $\varepsilon$ was fixed at $0.5$, whereas for kNN and related techniques, the number of neighbors $k$ was set to $3$, and the distance metric was defined as \textit{euclidean}~\cite{wasswa2025graph}.

\section{Results and Discussion}\label{results}
\begin{figure}[!ht]
     \centering
    \includegraphics[scale=0.8]{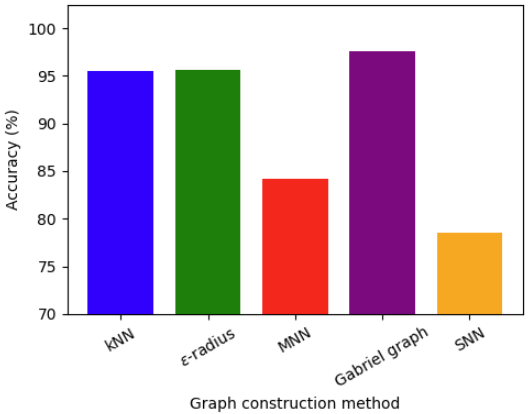}
    \caption{Classification accuracy comparison for the different graph data construction techniques}
    \label{fig:accuracy_comparison}
\end{figure}

The detection performance of the model was evaluated in terms of accuracy, precision, recall and F1-score. Figure~\ref{fig:accuracy_comparison} provides a visual comparison of the model's performance in terms of accuracy when the five different methods are used to transform the structured .csv $6$-dimensional dataset into a graph structure prior to model training. 

It is clear from the graph that converting the .csv dataset into a Gabriel graph structure results in better detection accuracy ($97.56\%$) while the SNN method leads to the lowest detection accuracy ($78.56\%$) followed by the mutual kNN method with a detection accuracy of $84.14\%$. On the other hand, the GAT models trained using the kNN and $\varepsilon$-radius graph construction methods achieved slightly lower detection accuracies of $95.54\%$ and $95.67\%$, respectively, compared to the model trained on the Gabriel graph, with neither model exhibiting a significant advantage over the other.

\begin{figure*}[!ht]
     \centering
     \hspace*{-0.2in}
    \includegraphics[scale=1.1]{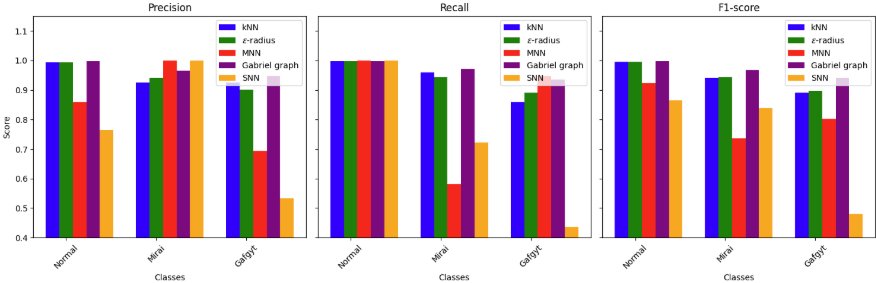}
    \caption{Performance comparison in terms of Precision, Recall and F1-score}
    \label{fig:per_class}
\end{figure*}

A similar pattern was observed in terms of other metrics including Precision, Recall and F1-score as shown in Figure~\ref{fig:per_class}. It is clear that the Gabriel graph consistently records a high detection performance for the three traffic classes across the three metrics. On the contrary, the SNN method, despite recording a high precision of $0.999$ for \enquote{Mirai} traffic and a high recall $0.999$ for \enquote{Normal} traffic, it records very low scores for other classes across the three performance metrics (for instance, the model records a precision of $0.534$, a recall of $0.436$ and an F1-score of $0.480$ for the \enquote{Gafgyt} attack family). 

The superior performance of the Gabriel graph can be possibly attributed to its construction principle, where edges are formed only if no other point lies within the hypersphere defined by two nodes. Applied to the 6-dimensional latent space of the N-BaIoT dataset, this preserves both local density and global separation of traffic patterns, aiding effective class discrimination.  In contrast, the SNN method underperformed as its reliance on shared neighbors likely fragmented the graph, failing to adequately connect diverse but related traffic instances. This led to poor class representation and imbalance across graph nodes, thereby reducing the model’s ability to generalize across all attack categories.

These findings clearly indicate that the graph construction method can significantly impact the effectiveness of GNN-based attack detection model.

\section{Conclusion}\label{conclusion}
In conclusion, GNN models are increasingly being integrated with attention mechanisms for effective detection of IoT botnets. Given the tabular nature of NetFlow data, five graph construction techniques---kNN, MNN, SNN, Gabriel Graph, and $\epsilon$-radius Graph---were examined following dimensionality reduction via a VAE. A GAT model was subsequently employed to classify network traffic in the N-BaIoT dataset as either benign or associated with one of the two IoT botnet malware families---Mirai or Gafgyt. Comparative results revealed that the Gabriel Graph achieved the highest classification accuracy of $97.56\%$, while the SNN method recorded the lowest performance at $78.56\%$. Further analysis of per-class instance recognition showed that the GAT model trained on the Gabriel Graph consistently achieved high scores across all traffic classes in terms of Precision, Recall, and F1-score. These findings underscore the crucial role of graph construction methods in enhancing GNN-based attack detection systems for IoT security.

\bibliographystyle{unsrtnat}
\bibliography{references}

\end{document}